\documentclass[twocolumn,aps,prl,showpacs]{revtex4}

\usepackage{epsfig}

\newcommand{\beq}{\begin{equation}}
\newcommand{\eeq}{\end{equation}}
\newcommand{\bea}{\begin{eqnarray}}
\newcommand{\eea}{\end{eqnarray}}

\begin{document}

\title{On the observation of the spin resonance in superconducting $CeCoIn_5$}
\author {A. V. Chubukov$^1$, L. P. Gor'kov$^2$, }
\affiliation{$^1$ Department of Physics, University of Wisconsin, Madison, WI
53706}
\affiliation{$^2$ National High Magnetic Field Lab. Florida State University, Tallahassee, FL}
\date{\today}
\begin{abstract}
Recent observation of a resonance
 spin excitation at $(1/2,1/2,1/2)$ in the superconducting state 
of $CeCoIn_5$ [C. Stock et al., Phys. Rev. Lett. {\bf 100} 087001 (2008)] was
 interpreted  as an evidence for $d_{x^2-y^2}$ gap symmetry,
 by analogy with the cuprates. This is true  
 if the resonance is a spin exciton. We argue that such description
 is undermined by the three-dimensionality of $CeCoIn_5$. We show
 that in 3D systems the excitonic resonance only emerges at strong coupling, and is weak. We argue in favor of the   alternative, magnon scenario, 
which does not require a $d_{x^2-y^2}$ gap.     \end{abstract}

\maketitle

Whether  Heavy Fermion materials (HF) and 
high temperature superconductors (HTSC) are governed by
 the same physics has been repeatedly discussed in the past. 
 Speculations about the similarities between the two 
 classes of materials started immediately  after Bednorz and 
Muller's discovery of the high-temperature superconductivity in the
 cuprates\cite{1}. To a large extent, the discussion at 
that time centered around an
 observation that in both systems spin degrees of freedom are among 
low-energy excitations, and superconductivity 
 should be unconventional if spins 
 are involved in the pairing.  It was later realized that, 
although electron-electron interactions and magnetism do play  
indispensable roles in both systems, the HTSC are 
different from HF in several important aspects. Among them: 
(i) the metallic state in the cuprates emerges as a result of a chemical 
doping into the antiferromagnetic (AFM) Mott insulator; (ii) the transport in the cuprates is very anisotropic, and the consensus is that the new physics is 
two-dimensional(2D) and uniquely related to $CuO_2$ planes; (iii) 
ARPES experiments in the optimally doped and underdoped cuprates 
do not see 
well-defined single particle excitations above $T_c$, except in the so-called 
nodal regions~\cite{2}.  On the contrary, in
 HF, like $UPt_3$ or $CeAl_3$,  
 the de Haas-van Alphen (dHvA)-type experiments clearly indicate that 
the physics is three-dimensional (3D), and that single-particle excitations 
 remain metallic and well-defined, although  electronic 
characteristics dramatically change upon lowering the temperature below 
an effective "Kondo temperature", $T_K$ , at which the local f-electron 
states "mix up" with the s-p-d bands.
 
The still unresolved issue is whether the difference in the effective dimensionality and the nature of normal state excitations
gives rise to a different symmetry of the superconducting order parameter in the HF compared to HTSC. There is a consensus among researchers that the
 SC order parameter in  HF is indeed non-$s-$wave, but there is no agreement 
 about its symmetry 
(see Ref.\onlinecite{3} for the summary of early results on HF ). 
The interest to this issue  resurfaced recently after the 
discovery of the  new class of HF with the chemical formula $CeMIn_5$ 
(labeled as $Ce115$), where 
$M=Co, Rh, Ir$ (Ref.\onlinecite{4}). 
These  materials are much closer to the cuprates  than other HF -- 
 they  crystallize into the layered tetragonal
 structure and possess a considerable anisotropy in their 
electronic properties. 
The dHvA experiments have found that at least one FS (electron 15-band ) 
has a pronounced quasi two-dimensional (Q2D) character~\cite{5,6}
 (see also Ref.\onlinecite{7}). The transport properties of $Ce115$s
 also display surprising similarities with Q2D cuprates~\cite{11}.

Another remarkable feature of the Ce115 materials,
also reminiscent of the cuprates,  is the  proximity 
between SC and AFM. The close interrelation between superconductivity 
and antiferromagnetism has been demonstrated by the experiments on  different alloys of $Ce115$s (Ref.\onlinecite{8}). 
The transition from SC to AFM can be 
tuned by applied pressure or the magnetic field (see Ref.\onlinecite{9} and references therein), or by substituting $Cd$  for $In$~\cite{10}. 

Such closeness between HTSC and $Ce115$s 
stimulated anew speculations that the superconductivity in both families of materials may have the same (magnetic) origin, and the gap symmetry is the same~\cite{9}. However, unlike HTSC, where the $d_{x^2-y^2}$ -symmetry of the SC gap is the well-established fact~\cite{2}, 
less is known about the gap  symmetry in HF $Ce115$s. 
In the absence of ARPES data, 
the  positions of nods in the SC gap for $Ce115$s  are subject of
  debates, and there are only indirect  arguments in favor of the  $d_{x^2-y^2}$ symmetry in $CeCoIn_5$~\cite{12}. Furthermore,
$CeCoIn_5$ is the multi-band superconductor, in distinction to the cuprates,
 and experiments show that some of its
 Fermi surfaces (FS) are prominently three-dimensional (3D)~\cite{5,6}. In 3D tetragonal materials,  there are more choices for the symmetry of the 
SC gap~\cite{VG}.

The subject of this paper is the analysis of recent
neutron experiments on $CeCoIn_5$ (Ref.\onlinecite{13}), particularly their ability to resolve the issue of the gap symmetry. The experiments observed 
 the spin resonance in $CeCoIn_5$ 
 at the commensurate AFM vector $Q^{3D}_0=(1/2,1/2,1/2)$, and at 
 temperatures below $1.35K$ (down to the lowest measured $T = 0.47 K$).
 The resonance is very likely related to superconductivity, as measurements 
above  $T_c =2.3K$, show only a shallow 
feature around $0.6meV$. 

The authors of \cite{13} used the analogy 
with the cuprates, where the spin 
resonance has been observed~\cite{14} at $Q^{2D}_0=( 1/2, 1/2)$, and argued that their data can be interpreted as the evidence that the superconducting gap in $CeCoIn_5$ is the same $d-$wave gap with $\Delta (p+Q^{3D}_0) =-\Delta (p)$, as in the cuprates. This is the case if the resonance is a spin exciton~\cite{exciton}.
 Below we demonstrate that three-dimensionality of $Q_0$,  at which the resonance has been observed in~\cite{13}, makes the analogy with the cuprates imprecise. 
  We argue that
 the resonance unlikely comes from 3D FS as we found that 
a 3D exciton is too weak. It may come from 
Q2D electron 15 band in $CeCoIn_5$, but then 
it should be visible not only at $(1/2,1/2,1/2)$, but also along the
 line $(1/2,1/2,b)$~ ($0<b<1/2$), which apparently is not the case experimentally. We argue that more likely explanation 
 is that the resonance is a magnon-type excitation of $f-$electrons, 
cleared by a superconductivity. 

To proceed, we first briefly review the situation in the 2D cuprates.
Several candidates for the resonance have been proposed, including 
 spin exciton, $\pi$-resonance, the mixture of the exciton and $\pi$-resonance,
 and a magnon, cleared up by a superconductivity~\cite{Macdonald}.
  We focus on the spin exciton scenario~\cite{exciton}
 because of its relation to $d_{x^2-y^2}$ superconductivity.   

The spin-exciton scenario associates the resonance with the feedback from
 the $d_{x^2-y^2}$ pairing on the spin susceptibility of itinerant fermions 
 near the AFM instability. The physics of this effect is
 not sensitive to the details of the model as long
 as the Fermi liquid concepts of the FS and the electron-hole excitations
 around it are preserved. The full dynamic spin susceptibility of itinerant
 fermions can be quite generally expressed as
\begin{equation}
\chi(Q, \Omega) = \mu^2_B \frac{\chi_0}{\xi^{-2} - \Pi (Q, \Omega)}
\label{1}
\end{equation}
where $\mu_B$ is Bohr magneton, and dimensionless 
$\xi$ is proportional to the magnetic correlation length ($\xi^{-1} =0$
 signals the onset of AFM order), and $\Pi (Q, \Omega)$ is the polarization operator which has the form
\begin{eqnarray}
&&\Pi (Q,\Omega)= 16 g^2 \chi_0 T \sum_{\omega} \int \frac{d^3p}{(2\pi)^3} 
\nonumber \\
&& \left(G(p,\omega)G(p+Q,\omega +\Omega)+ F(p,\omega)F^{+}(p+Q,\omega+\Omega)\right)\nonumber \\
\label{eq_2}
\end{eqnarray}
where $G$, $F$, and $F^{+}$ are the normal
 and anomalous Gor'kov Green functions, correspondingly, and 
 $g$  is the exchange constant for the interaction between 
conduction electrons and their collective spin excitations.
Spin-exciton scenario assumes
  that the FS crosses the magnetic Brillouin zone boundary at eight points (hot spots) separated by 
$Q = Q^{2D}_0$ [the prefactor $16$ in (\ref{eq_2}) is $2*8$, where 
extra $2$ is the spin factor]~\cite{17}. It further assumes that
 spin excitations with momenta near $Q^{2D}_0$ 
are completely overdamped in the normal state 
due to strong Landau damping into particle-hole fermionic pairs, such that
 $Im \Pi_n (Q, \Omega) \approx i\gamma \Omega$, where $\gamma$ is proportional
 to $g^2 \chi_0 \nu (E_F)/E_F$, where $\nu(E_F)$ is the density of states at the FS (note that $g^2 \chi_0 \nu (E_F)$ is dimensionless)  
The real part of $\Pi_n (Q, \Omega)$ 
 in the normal state can be safely approximated by its  static value 
$\Pi_n (Q,0)$ which we absorb into $\xi^{-2}$. In the superconducting state, 
fermions at hot spots become gapped, and the behavior of $\Pi (Q, \Omega)$ changes. Expanding near the FS and 
integrating in  Eq. (\ref{eq_2}) over the transverse momentum 
component, we obtain~\cite{exciton}
\beq 
\Pi_{sc} (Q, \Omega) = \frac{i \gamma}{2} \int d\omega
 \left[1- \frac{\omega_+ \omega_-  + \Delta_1 \Delta_2}{\sqrt{\omega^2_+ 
- \Delta_1^2} \sqrt{\omega^2_- - \Delta_2^2}}\right] 
\label{4}
\eeq 
where 
$\omega_{\pm} = \omega \pm \Omega/2 + i\delta {\text sign} (\omega \pm \Omega/2)$, and 
$\Delta_1$ and $\Delta_2$ are the gaps at the two 
hot spots separated by $Q$. 

The remarkable feature in 2D that distinguishes between $s-$wave, 
$d_{xy}$ and $d_{x^2-y^2}$ pairings
comes about from the factor $\Delta_1 \Delta_2$
 in (\ref{4}): if this factor is negative, which is the case for
 the $d_{x^2-y^2}$ 
 symmetry, but not the other two symmetries, then (i) $\Pi (Q,0)=0$, i.e., the value of $\xi$ is unaffected by superconductivity, 
 and (ii)  $Re \Pi_{sc} (Q,\Omega)$ 
diverges logarithmically at $ 2\Delta_1$, 
as $\gamma \Delta_1 \log {\Delta_1/|2\Delta_1 -\Omega|}$, and 
$Im \Pi_{sc} (Q,\omega)$ jumps discontinuously at $\Omega =2\Delta_1$ 
from zero to $\pi \gamma \Delta_1$.  At $\Omega <2\Delta_1$, 
$Im \Pi_{sc} (Q,\Omega) =0$ while $Re \Pi_{sc} (Q,\Omega)$ gradually decreases
and behaves as  $\pi \gamma \Omega^2/(8 \Delta_1)$ at small $\Omega$. 
This behavior of $ \Pi_{sc} (Q,\Omega)$ guarantees 
that $\chi(Q,\Omega)$ from (\ref{1}) has a pole somewhere 
below $2\Delta_1$ (a spin exciton)~\cite{comm_a}.
 The exciton frequency is close to $2\Delta_1$ at small $\xi$, but progressively shifts down as 
$\xi$ increases.
At large $\xi$, the pole is at $\Omega =(8 \Delta_{1}/(\pi\gamma))^{1/2}  \xi^{-1}$. An important fingerprint of the 2D spin exciton 
 scenario is a negative momentum 
dispersion of the peak, which originates from the fact that
the energy of the exciton must vanish at the momentum 
 which connects nodal points on the Fermi surface

For $s-$wave gap symmetry, $Re \Pi_{sc} (Q, 0)$ is negative, $Re \Pi_{sc} (Q, 2\Delta_1)$ does not diverge, and the calculations show~\cite{ctj} that $Re \Pi_{sc} (Q, \Omega)$ remains negative for all $\Omega$ in which case spin exciton does not emerge.      

Consider now whether one can pass this consideration to a three-dimensional 
$CeCoIn_5$.
The dHvA experiments on $Ce115$ have found several small 3D FS
 and a large 3D hole FS, whose size is large enough such that this FS
 contains  hot lines -- contours of FS
 points connected by a 3D diagonal $Q^{3D}_0$. 
To verify whether such FS may be responsible for the spin resonance, we now re-evaluate the staggered spin susceptibility, Eq. (\ref{1}), for
  a model system with a spherical FS with a diameter $r_0>\sqrt{3}/4$.
The  hot lines on this FS  are specified 
by $\cos \theta + \sqrt{2} \sin \theta \sin (\phi + \pi/4) = 3/(4r_0)$, where
$\theta$ and $\phi$ are azimuthal and polar angles for a point on a hot line.
We considered various symmetries of the pairing gap originating from 
 different representations of  the tetragonal group, $D_4$~\cite{VG}:  (i) 
 one-dimensional 
representations $A_{1g}$  ($\Delta_k \propto  k_x^2 + k_y^2$), $B_{1g}$  ($\Delta_k \propto  k_x^2-k_y^2$), $B_{2g}$ ($\Delta_k \propto  k_x k_y$), $A_{2g}$ ($\Delta_k 
\propto  k_x k_y (k_x^2-k_y^2)$) (the $A_{2g}$ gap would have too many nodes), and (ii)  two-dimensional representation $E_g$  
($\Delta \propto k_z(k_x+i k_y), ~\Delta \propto k_zk_x$, and $\Delta \propto 
k_z(k_x+k_y)$). We found 
 that the 3D resonance in $\chi (Q, \Omega)$ is still possible only
 if the pairing gap retains the same $d-$wave, $k^2_x-k^2_y$ 
 symmetry as in the cuprates, i.e., at the spherical FS
 $\Delta_k = \Delta \sin^2\theta \cos 2\phi$
 (in 3D notations, $\Delta \propto Y_{2,2} + Y_{2,-2}$, where $Y_{l,m} (\theta, \phi)$ are spherical harmonics). 

Compared to 2D case,  the calculation of $\Pi (Q,\Omega)$ using 3D
 version of Eq. (\ref{4}) involves a summation along the whole hot line. 
In the normal state, this does not lead to a new physics -- we still 
have Landau overdamped excitations with 
$\Pi (Q, \Omega) = i \gamma_{3D} \Omega$.  
In a superconducting state, however, the integration along
 hot lines leads to three essential 
differences with the 2D case:
\begin{itemize}
\item
1. by symmetry,
 hot lines necessary cross the directions $\phi = (3\pi/4)(2n+1)$, $n =0,1,2,3$, where the $d-$wave gap vanishes. This implies that $Im \Pi (Q,\Omega)$ is finite at any non-zero frequency
\item
2. for an arbitrary point $k_F$ along a hot line, the gaps $\Delta_{k}$ and $\Delta_{k+Q}$ are not simply related, the condition $\Delta_k = - 
\Delta_{k+Q}$ is satisfied only for special symmetry 
points along a hot line. As a consequence, 
$Re \Pi (Q,0) \propto \oint d{\bf k} (\Delta_{k} + \Delta_{k+Q})^2$ at $Q = Q^{3D}_0$ is finite,
 negative, and of the order of $\gamma_{3D} \Delta$
\item
3. the logarithmic divergence of $Re \Pi (q, \Omega)$ at a 3D analog of 
$2\Delta_1$ is removed due to gap variation along a hot line. 
A simple calculation shows that $Re \Pi (q, \Omega)$ only has a cusp at $\Omega \sim 1.55 \Delta$. 
\end{itemize}
\begin{figure}[h]
\epsfig{file=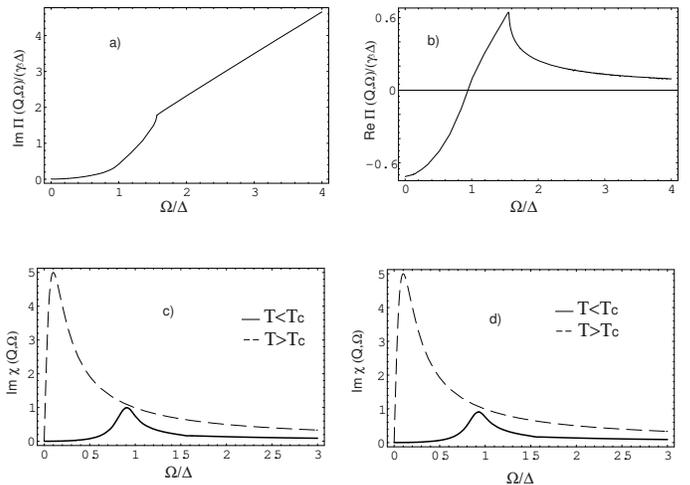,width=9.cm}
\caption{Panels a) and b) --  real and imaginary parts of the polarization operator
 $\Pi (Q, \Omega)$ for 3D spherical FS with hot lines. At
 large frequencies, $Im \Pi (\Omega) = \Omega/\Delta$, where $\Delta$ is the maximum of the $d_{x^2-y^2}$ gap.  We set $r_0 =1$. 
 Panels c) and d) -- the imaginary part of the full spin susceptibility, Eq. (\protect\ref{1}) (in arb. units) in the normal state (dashed) and $d-$wave superconducting state (solid)  
for $\xi^{-2}/(\gamma_{3D} \Delta) = 0.1$ (c), and $\xi^{-2}/(\gamma_{3D} \Delta) = 
0.25$ (d).
Observe that $Im \chi (Q, \Omega)$ in a $d_{x^2-y^2}$ superconductor 
 does has a ``resonance'' peak, but its intensity is not enhanced compared 
 to the normal state.} \label{fig1}
\end{figure}
 
In Fig.\ref{fig1} we plot real and imaginary parts of the polarization operator, and the full $Im \chi (Q, \Omega)$, Eq. (\ref{1}) for two different $\xi$.
 We see that the resonance is still there, but it is rather broad, and the intensity is not  enhanced compared to that of the normal state.
This is in contrast to the sharp peak observed in the neutron experiment. 
Likely then, the observed peak does not come from 3D Fermi surfaces.

Still, the observed resonance could potentially be a spin exciton. As we said, 
 dHvA experiments have found, that $Ce115$ has at least one 
 Q2D FS (a FS for the electron 15-band). 
The last FS is a slightly corrugated parallelepiped 
with the base close to  a 
square. If this FS was strictly cylindrical along the z-axis, the hot lines would be
 parallel to $z$, and for $d_{x^2-y^2}$  symmetry of the gap, 
$\Delta_{k+Q} =-\Delta_k$ for any $k-$point along a hot line. The integration over $z$ then would be harmeless, and the excitonic resonance would be identical to the one in a purely 2D system, where it is sharp.
There are two requirements: one is that the resonance frequency must be smaller than twice the gap maximum.  The resonance in $CeCoIn_5$ 
 is observed at $\Omega_{res} =0.6 meV$, below $T_c =2.3K$. 
If we use a conservative estimate $2\Delta_{max} \sim 4T_c$, we do find 
 that $\Omega_{res}$ is indeed smaller than $2\Delta_{max}$. 
 The other requirement is that 
such Q2D FS must be large enough to contain hot spots in the 
$xy$ plane. For this,  
 the area encircled by the extreme dHvA orbits, $S_{ext}$, (marked as 
$\alpha_{1,2,3}$ in Refs.\cite{5, 6}) must exceed a quarter of the area of the 
Brillouin Zone: $S_{BZ} = (2\pi/a)^2$ ($a = 4.61410^{-8}cm$ in $CeCoIn_5$).
One has:
\begin{equation}
 S_{ext}= {\cal F} \left(\frac{2e\pi}{c{\hbar}}\right),
\end{equation}
where ${\cal F}$ is the experimental dHvA frequency. 
 According to Refs. (\cite{5}-\cite{7}), the experimental ratio $4S_{ext}/S_{BZ}$
 for $\alpha_1$ orbit is $1.15 >1$. This implies that the area of 
a Q2D FS is indeed large enough to contain hot spots.

Such explanation of the resonance in $CeCoIn_5$, however, 
 disagrees with the experiments in one important aspect.
 If the resonance is an effective 2D exciton,
 it should be present not only for 
$(1/2,1/2,1/2)$, but also for the whole set of momenta $Q = (1/2,1/2,b)$,
 where $0<b<1/2$. In~\cite{13} the resonance 
intensity is clearly peaked at $b=1/2$.  The absence of strong 
resonances at other momenta  
may be the consequence of the facts that the Q2D FS is not a perfect cylinder as evidenced by the existence of 3 different dHvA orbits, 
that $\Delta$ generally varies  along $z$ axis, and  that 3D FS 
 also contribute to $\Pi (Q, \Omega)$

Still, the constraints on the excitonic description 
call for another possible explanations  for the resonance in $CeCoIn_5$.
 A potential candidate  is the ``magnon'' scenario, originally suggested for the cuprates~\cite{15}. 
In application to $CeCoIn_5$, this scenario 
 assumes that this system is close to a AFM state,
 and contains quasi-localized spins of $f-$electrons 
coupled to each other and to  conduction electrons. To account for the localized spins, one should include into the denominator of Eq. (\ref{1}) an
 additional  term  $(\Omega/\omega_0)^2$  which in the hypothetical absence of the damping would give rise to a magnon mode with $\Omega = \omega_0 \xi^{-1}$.
 For itinerant fermions, $\omega_0$ is of order Fermi energy,
 and such term is negligible for $\omega \sim \Delta$. For HF materials, $\omega_0$ is much smaller, and  a magnon mode well may have energy comparable to $\Delta$. 

This mode should in principle
 be present both above and below $T_c$, but  in the normal state  it is washed out by Landau damping. 
In a superconductor, Landau damping is strongly reduced at 
 $\Omega \leq 2\Delta$, and the magnon may ``clear up''. 
Furthermore, in 3D, $Re \Pi (Q, \Omega)$ is negative at $Q = Q^{3D}_0$, what 
 effectively reduces $\xi$ and hence  increases  the stability of SC state 
against the onset of AFM. 

Parameterwise, a magnon is overdamped above $T_c$ if 
 $\gamma_{3D} \geq \xi^{-1}/\omega_0 \sim \xi^{-2}/\Delta$. Using 
$\gamma_{3D} \sim g^2 \chi_0 \nu (E_F) /E_F$, $g^2 \chi_0 \nu (E_F) \sim 1$,
 and taking $E_F \sim T_K$ for HF materials ($T_K$ is a Kondo temperature), 
we find that the inequality amounts to $(\Delta/T_K) \geq \xi^{-2}$, which 
 can well be satisfied in $Ce115$ due to its proximity to AFM.
  Below $T_c$, the damping is reduced and 
$Re \Pi (Q, \Omega)$ appears instead. 
This $Re \Pi$ scales with frequency as $\gamma_{3D} \Omega^2/\Delta$, such that  the denominator of Eq. (\ref{1}) contains two $\Omega^2$ terms, and  becomes (roughly) 
\begin{equation}
 \xi^{-2}_{eff} - \Omega^2 \left(\frac{1}{\omega^2_0} + \frac{\gamma_{3D}}{\Delta}\right),
\end{equation}
 where $\xi^{-2}_{eff} = \xi^{-2} - Re \Pi (0,0)$. Using $\omega_0 \sim \Delta \xi$ and $\gamma_{3D} \geq \xi^{-2}/\Delta$, we see that the two $\Omega^2$ contributions are of comparable magnitude, i.e., the restored resonance is partly a magnon, partly an exciton. We also note that,
 for the model considered above,
 $Re \Pi (Q, \Omega)$ changes sign and is quite small 
at $\Omega \sim \Delta$ (see Fig. 1b). 
 If the magnon energy is in this range, the 
   feedback electronic contribution is negligible,
 and the resonance below $T_c$ is simply a
 magnon, restored by superconductivity. 

Quite obviously, the magnon scenario
 is not peculiar to $d_{x^2-y^2}$ gap symmetry, and holds also
 if the superconducting gap has an $s-$wave symmetry.
The only difference is that, in the 
$s-$wave case,  $Re \Pi (Q,0)$ does not change sign for $Q = Q^{3D}_0$, and 
remains negative for  all frequencies~\cite{ctj}.

A possible way to distinguish between magnon and spin-exciton scenario is to study the dispersion of the resonance peak, $\Omega(Q \neq Q^{3D}_0)$. In the spin-
exciton scenario, the dispersion must be negative, at least in some range of $Q$, because once $Q$ becomes equal to the diameter of the FS (modulo $2\pi$), 
it must, by symmetry, connect the points on the FS for which $d_{x^2-y^2}$ gap vanishes.
 For such $Q$, the frequency of the exciton must vanish. On the other hand, if the resonance is a magnon, its frequency should 
 monotonically increase with  deviations from $Q^{3D}_0$.   

To summarize, in this paper we addressed the issue whether the resonance neutron peak observed in the superconducting state of $CeCoIn_5$ may be 
a spin exciton peculiar to $d_{x^2 -y^2}$ symmetry of the pairing gap.  
We argued that the peak is unlikely a 3D exciton originating from 3D FS of $CeCoIn_5$, as such peak is too weak and 
does not raise above the normal state result. The sharp excitonic resonance still  might come from Q2D FS of  $CeCoIn_5$, but such resonance
 should be observed not only at $Q^{3D}_0 = (1/2,1/2,1/2)$ but along the whole line of  $(1/2,1/2,b)$.  We argued that the  experimental data~\cite{13} do not
 immediately support this scenario. We presented
 another, more plausible explanation of the  neutron data, namely 
that the resonance is a  magnon-type excitation of localized $f-$electrons,
 restored and modified  by superconductivity. If the resonance is a magnon, its observation is not an argument for $d_{x^2-y^2}$ gap symmetry.
  We suggest to study the dispersion of the resonance to 
  distinguish between the two scenarios.

 We are thankful to 
Z. Fisk for the discussion and for 
attracting our attention to Ref. \cite{13}.
AVC was supported by NSF-DMR 0604406. 
The work of LPG was supported by the NHMFL through NSF Cooperative Agreement No. DMR-9527035 and the
State of Florida.


\begin{thebibliography}{99}
\bibitem{comm_a} The logarithmic divergence of 
$Re \Pi_{sc} (\omega)$ at $\Omega =2\Delta_1$ is smoothed
 at a finite T or in the presence of an impurity scattering,
 and the resonance is actually only observable if $\xi^{-2}/(\gamma\Delta_1)$ 
 is large enough.
\end{thebibliography}
\end{document}